\documentclass[twocolumn]{aastex631}
\usepackage{showyourwork}
\usepackage{amsmath,esint}
\usepackage{hyperref}    
\usepackage{fontawesome} 

\begin{document}

\title{A General, Differentiable Transit Model for Ellipsoidal Occulters: Derivation, Application, and Forecast of Planetary Oblateness and Obliquity Constraints with JWST}

\author{\href{https://orcid.org/0000-0001-9145-8444}{Shashank Dholakia}} 
\affiliation{School of Mathematics and Physics, The University of Queensland, St Lucia, QLD 4072, Australia}

\author{\href{https://orcid.org/0000-0001-6263-4437}{Shishir Dholakia}}
\affiliation{Centre for Astrophysics, University of Southern Queensland, West Street, Toowoomba, QLD 4350, Australia}

\author{\href{https://orcid.org/0000-0003-2595-9114}{Benjamin J. S. Pope}}
\affiliation{School of Mathematics and Physics, The University of Queensland, St Lucia, QLD 4072, Australia}
\affiliation{Centre for Astrophysics, University of Southern Queensland, West Street, Toowoomba, QLD 4350, Australia}

\begin{abstract}
Increasingly precise space-based photometry uncovers higher-order effects in transits, eclipses and phase curves which can be used to characterize exoplanets in novel ways. The subtle signature induced by a rotationally deformed exoplanet is determined by the planet's oblateness and rotational obliquity, which provide a wealth of information about a planet's formation, internal structure, and dynamical history. However, these quantities are often strongly degenerate and require sophisticated methods to convincingly constrain. We develop a new semi-analytic model for an ellipsoidal object occulting a spherical body with arbitrary surface maps expressed in terms of spherical harmonics. We implement this model in an open-source \textsc{Jax}-based Python package \href{https://github.com/shishirdholakia/eclipsoid}{\texttt{eclipsoid}}, allowing just-in-time compilation and automatic differentiation. We then estimate the precision obtainable with JWST observations of the long period planet population and demonstrate the best current candidates for studies of oblateness and obliquity. We test our method on the JWST NIRSpec transit of the inflated warm Neptune WASP-107~b and place an upper bound on its projected oblateness $f<0.23$, which corresponds to a rotation period of $P_{\mathrm{rot}}>13$\,h if the planet is not inclined to our line of sight. Further studies of long-period exoplanets will necessitate discarding the assumption of planets as spherical bodies. \texttt{Eclipsoid} provides a general framework allowing rotational deformation to be modelled in transits, occultations, phase curves, transmission spectra and more. \href{https://github.com/shishirdholakia/eclipsoid/tree/main}{\faGithub} \href{https://github.com/shashankdholakia/oblate-planets-paper/tree/main}{\faBarChart}
\end{abstract}

\section{Introduction} \label{sec:intro}
The contemporary study of exoplanets frequently assumes that both the star and planet are spherical bodies. While this is a close approximation to the truth, the deviations from this case are informative and contain valuable information which may be discarded by assuming sphericity. Two dominant deviations from sphericity are expected for certain exoplanets: oblateness is an elongation towards the equator due to the impact of rapid rotation. The resulting shape, an oblate spheroid, can be parametrized by rotating an ellipse around the \textit{minor} axis, which in this case the axis of the body's rotation. Prolateness occurs due to the effect of tides; a prolate spheroid can similarly be parametrized by rotating an ellipse around the \textit{major} axis, in this case the tidal axis.

Planetary oblateness has never been measured unambiguously outside the Solar System, but is important for some planets that orbit at wide enough separations from their host stars that tides do not synchronize the rotation to the orbital period of the planet \citep{seager2002a}. For these planets, rapid rotation results in deformation due to the centrifugal force causing deviation from spherically symmetric hydrostatic equilibrium. This results in a planet shaped as an oblate spheroid with a larger radius at the equator than the poles. Saturn in the solar system rotates with a rotation period of $P_{\rm rot} \approx $ 10\,hours and 39\,minutes, resulting in an oblateness factor of $f_o=0.098$ \citep{Saturn}, being the ratio of the equatorial radius to the polar radius \citep{davies1980}. Jupiter rotates in roughly 10 hours, but being more massive is less distorted, at about $f_o=0.065$ \citep{Jupiter}. Uranus and Neptune are slower rotators, less-oblate at $f_o<0.03$ \citep{Uranus,Neptune}. 

The relationship between the rotation rate and oblateness is given to first order by 
\begin{equation}
    P_{\mathrm{rot}} = 2 \pi \sqrt{\frac{r_{eq}^3}{2 G m}\frac{1}{f_o}}
\end{equation}

\noindent where $P_{\mathrm{rot}}$ is the rotation period, $r_{eq}$ the equatorial radius, $G$ Newton's constant, and $m$ the mass of the body \citep{chandrasekhar1969, seager2002b}.

When a non-negligible portion of the mass of the planet is distended at the equator, the gravitational potential must be expanded to higher order to obtain a more precise formula \citep{carter2010a}, 

\begin{equation}
    P_{\mathrm{rot}} = 2 \pi \sqrt{\frac{r_{eq}^3}{G m}\frac{1}{\left(2 f_o - 3 J_2\right)}}
\end{equation}
where the $J_2$ moment is the 2nd-order coefficient from the expansion of the gravitational potential of the planet into spherical harmonics. 

Also relevant to the study of planetary rotation is the planetary rotational obliquity; the angle between the rotational and orbital angular momentum vectors for a planet. For the giant planets in the Solar System, Uranus is most misaligned with a tilt of $98^\circ$, followed by Neptune at $30^\circ$, Saturn at $27^\circ$ and Jupiter, which is nearly aligned at $3^\circ$. These numbers imply that rotational alignment is not necessarily common, even for a relatively quiescent evolution through a protoplanetary disk, and can be due to interactions with other planets in the system. Like oblateness, this effect has not been detected unambiguously for a transiting exoplanet. Measuring this effect could be necessary to explain why certain planets are tidally inflated \citep{millholland2019}, or probe the dynamical history of an exoplanet system \citep{millholland2024}. 

The advent of high-precision space-based photometry has enabled planetary parameters to be constrained precisely for a large number of transiting planets. Previous works have attempted to quantify the prospects to detect oblateness and constrain obliquities with \textit{Kepler} such as \citep{zhu2014} and \textit{Spitzer} \citep{carter2010a}. Most recently, the \textit{James Webb Space Telescope} (JWST) has demonstrated unprecedented white light-curve photometric precision, as low as 50\,ppm \citep{wasp39b_main}. 

The effect of rotational deformation and obliquity can be measured in precise transit photometry by modeling the subtle effects on the transit shape. A body that is rotationally (and/or tidally) deformed will take the shape of an ellipsoid in three dimensions, but during an occultation is projected into an ellipse. Importantly, the transit method is sensitive only to the \textit{projected} oblateness and obliquity (throughout this manuscript we use $f$ and $\theta$ to refer to these projected quantities), as opposed to the true oblateness $f_o$ and obliquity $\theta_{\mathrm{o}}$, which are projected by the planet's inclination angle with respect to the observer. Furthermore, to first order, the transit method is only sensitive to the \emph{area} occulted, as opposed to the true shape of the occulter. There is then a spherical planet with an equivalent area, with radius

\begin{equation} \label{eq:r_circ}
    r_{\mathrm{circ}} = \frac{r_{eq}}{\sqrt{1 - f}}
\end{equation}
where $r_{eq}$ is the ellipsoidal occultor's projected equatorial radius. Such a planet would have the same transit depth and could fit a low-precision observation of a transit \citep{seager2002a}. Likewise, a continuum of oblateness values with an equivalent area would also have the same transit depth and roughly the same transit shape. The majority of the information on oblateness and obliquity is therefore in the ingress and egress of the transit. The degeneracy between planets of various oblateness values in realistic light curves necessitates a careful model to understand the full set of oblateness and obliquity values consistent with a given dataset.

In Section~\ref{sec:model}, we describe a new forward model for light curves of spherical bodies with ellipsoidal occulters. We then describe the implementation of this model in our new open-source Python package \texttt{eclipsoid} in Section~\ref{sec:eclipsoid}. The package is built using the \textsc{Jax} framework \citep{jax}, which allows automatic differentiation of the model with respect to its parameters. We describe a use of this in Section~\ref{sec:jwstdetect}, where we perform an injection-recovery exercise with Hamiltonian Monte Carlo to quantify the detectability of ellipsoidal exoplanets using JWST and other instruments. Lastly, in Section~\ref{sec:lctest}, we apply the new model to the JWST/NIRSpec white light curve of WASP-107~b \citep{Sing2024}, an inflated sub-Saturn and demonstrate that we can constrain the oblateness and prolateless of the exoplanet. 

During the writing of this manuscript we became aware of the following similar packages that aim to model occultations of ellipsoidal objects: \texttt{JoJo} \citep{liu2024} and \texttt{squishyplanet} \citep{squishyplanet}. These models are similar to ours in several respects, including the use of Green's theorem to reduce the computational cost of evaluating the flux integrals, and in the case of \texttt{squishyplanet}, the use of the \textsc{Jax} framework. Our method is more general than these in that it allows arbitrary surface maps on either the star and planet, which may be useful for modelling full phase curves as well as marginalizing over stellar spots. In addition, our method \texttt{eclipsoid} is designed to be fully differentiable up to second order, allowing the use of gradient-based techniques such as HMC and variational inference, which scale better when the number of parameters is large.

We are committed to open science: this paper is produced using public data and open-source software, with reproducible calculations. At the end of every figure caption, we link to a GitHub repository containing a Python script used to generate the figure. The full set of scripts and notebooks involved in the analysis in this work can be found at the following link as a \texttt{showyourwork} repository \citep{luger2021}: \href{https://github.com/shashankdholakia/oblate-planets-paper}{\faGithub}

\section{Ellipsoidal Model}
\label{sec:model}
The transit model we use is based on the formalism of \texttt{starry}  \citep{starry2019}, which provides closed-form expressions for occultations involving circular bodies in projection with general surface maps expressed in terms of spherical harmonics. \citet{dholakia2022} described an extension of the \texttt{starry} framework for an oblate occulted body, such as a rapidly rotating star. Here we present an extension for the case of any occulter which is an ellipse in projection. 


We summarize the method here, and refer readers to the full text for the details. 
Our goal is to compute the flux observed during an occultation of a star by a body, which can be written as a surface integral of the intensity over the visible region of the star as

\begin{equation}
   F = \oiint\limits_{\mathrm{S}(x,y)} I(x,y) \ dS
\end{equation}
where the surface $S$ parametrizes the unobscured portion of the stellar disk and $I$ is the specific intensity at a point $(x,y)$ on the projected surface of the star.

We start with a vector representing the stellar intensity in terms of spherical harmonics $\mathbf{y}$, and transform it into Green's basis as in Equation~21 of \citet{starry2019}. We then have

\begin{equation}
   F = \oiint\limits_{\mathrm{S}(x,y)} \mathbf{\tilde{g}}^\mathsf{T}(x,y) \ \mathbf{A}\ \mathbf{R}\ \mathbf{y}\ dS 
\end{equation}
where the vector $\mathbf{y}$, the rotation matrix $\mathbf{R}$ and the change of basis matrix $\mathbf{A}$ are not dependant on $x$ and $y$ and can consequently be pulled out of the integral, leaving only the Green's basis in the surface integral. Using the vector function $\mathbf{G_n}$, defined as the anti-exterior derivative of the $n$th term in Green's basis, we can write

\begin{equation} \label{eq:greensintegral}
   \oiint\limits_{\mathrm{S}(x,y)} \mathbf{\tilde{g}}_n(x,y)\ dS \\
   = \oint \mathbf{G}_n(x,y) \cdot d\mathbf{r}
\end{equation}
where $\mathbf{r}$ is a vector function along the closed boundary of the region $S(x,y)$. We can then further decompose the integral in Equation~\ref{eq:greensintegral} into a section along the stellar projected disk and the occulter's projected disk (see Figure~\ref{fig:integral_bounds}):

\begin{equation} \label{eq:pandq}
    \oint \mathbf{G}(x,y) \cdot d\mathbf{r} = \mathcal{Q}(\mathbf{G}_n) - \mathcal{P}(\mathbf{G}_n)
\end{equation}

From here, we deviate from the \texttt{starry} framework to solve the line integrals around the star and elliptical occulter. First, we apply a rotation by an angle $\theta$ into a frame where the occulter's major axis is aligned with the x-axis. We then must solve for the points of intersection between the star and the occulter. 

\subsection{Integration Bounds}

First, it helps to consider the circular case, where the star is parametrized as the unit circle
\begin{equation} \label{eq:unitcircle}
x^2 + y^2 = 1
\end{equation}
and the occulter as an off-center circle with radius $r_{o}$ (we will use subscript $o$ throughout to denote the occulter, for full generality, agnostic to whether it is a star or planet) as
\begin{equation} \label{eq:circularplanet}
(x-x_o)^2-(y-y_o)^2 = r_o^2
\end{equation}

\noindent We can solve for the intersection points by solving for all $(x,y)$ which satisfy both equations, which yields a quadratic equation with either 0, 1 or 2 real solutions.

In the elliptical case, we can modify Equation~\ref{eq:circularplanet} by deforming the occulter along the y axis by a value $b$, now taking the radius $r_{eq}$ to represent the projected equatorial radius of the occulter:
\begin{equation} \label{eq:ellipticalplanet}
(x-x_{o})^2-\frac{(y-y_o)^2}{b^2} = r_{eq}^2
\end{equation}
where we define $b=1-f$. Here we emphasize that, henceforth, $r_{eq}$ and $f$ refer to the \textit{projected} equatorial radius and oblateness respectively. 
Solving for y in the above equation and then plugging it into Equation~\ref{eq:unitcircle} yields a quartic polynomial of the form: 
\begin{equation} \label{eq:quarticform}
Ax^4 + Bx^3 + Cx^2 + Dx + E = 0
\end{equation} 
where
\begin{equation} \label{eq:quarticcoeffs}
\begin{aligned}
A &= \frac{b^4 - 2b^2 + 1}{4y_o^2}\\
B &= \frac{-b^4x_o + b^2x_o}{y_o^2}\\
C &= \frac{-b^4r_{eq}^2 + 3b^4x_o^2 + b^2r_{eq}^2 - b^2x_o^2 + b^2y_o^2 + b^2 + y_o^2 - 1}{2y_o^2} \\
D &= \frac{b^4r_{eq}^2x_o - b^4x_o^3 - b^2x_oy_o^2 - b^2x_o}{y_o^2} \\
E &= \frac{1}{4y_o^2} \cdot (b^4r_{eq}^4 - 2b^4r_{eq}^2x_o^2 + b^4x_o^4 - 2b^2r_{eq}^2y_o^2 \\& \hspace{4em} - 2b^2r_{eq}^2 + 2b^2x_o^2y_o^2 + 2b^2x_o^2 + y_o^4 - 2y_o^2 + 1)\\
\end{aligned}
\end{equation}

We note that other methods of finding the flux in transit of ellipses in projection also require solving for quartic polynomials \citep[i.e exorings in][]{rein2023} or shells of intensity on exoplanets \citep{luger2017}. We solve the roots of this polynomial using eigendecomposition of the companion matrix (see Section~\ref{sec:eclipsoid} for details of the implementation). The solution gives the $x$ values of the intersection points where the projected disk of the star and the occulter coincide. We can then find the corresponding $y$ values by plugging it back into the formula for either the occulter or occulted body. 

\subsection{Star Boundary Integral}
\begin{figure}[ht!]
    \script{integral_bounds.py}
    \begin{centering}
        \includegraphics[width=\linewidth]{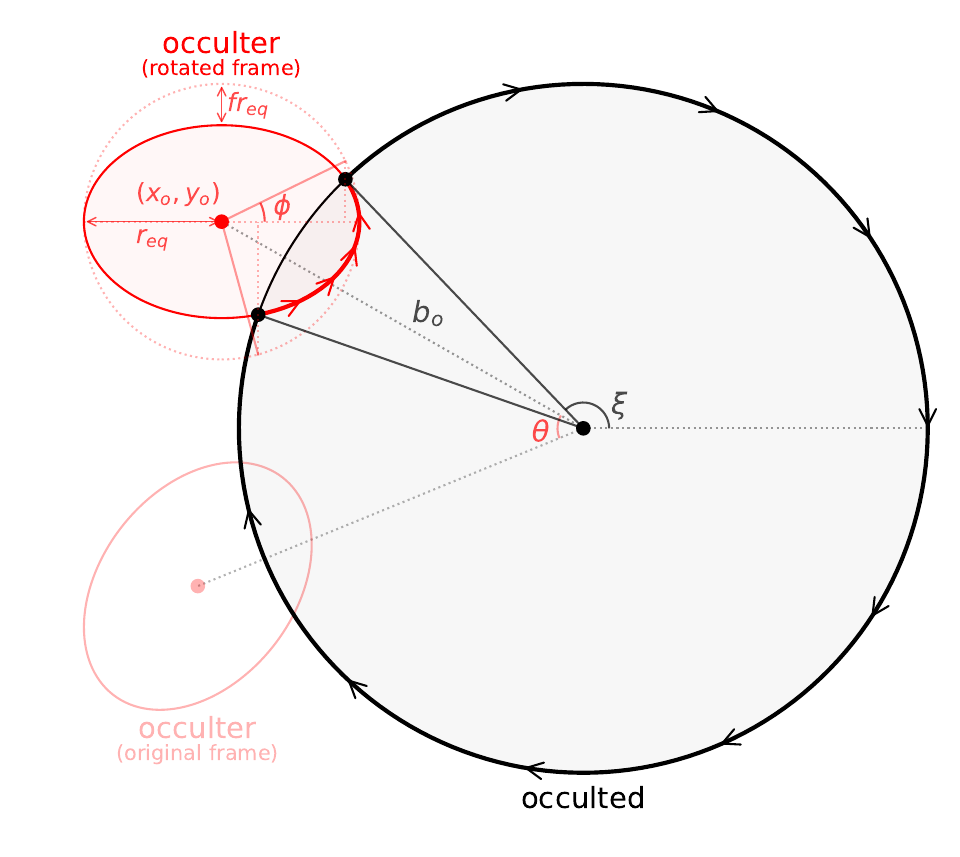}
        \caption{Geometry of the problem of computing the flux due to an oblate occulter as presented in this paper. The true sky frame is rotated to a standard form, with the occulting ellipse centred at $(x_o,y_o)$, and the contour integral is taken along the elliptical arc inside the occulted circle. \href{https://github.com/shashankdholakia/oblate-planets-paper/blob/main/src/scripts/integral_bounds.py}{\faGithub}
        }
        \label{fig:integral_bounds}
    \end{centering}
\end{figure}

The first integral $\mathcal{Q}(\mathbf{G}_n)$ is performed around the boundary of the occulted body's projected disk (bolded black border in Figure~\ref{fig:integral_bounds}.) While the integrand is the same as in \citet{starry2019}, the bounds of the integral are the roots of the quartic polynomial shown in Equation~\ref{eq:quarticcoeffs}. We compute the angle $\xi$, defined as the angle between the x-axis and a given intersection point, for all the intersection points. We then sort these angles in clockwise order. 

\subsection{Occulter Boundary Integral}
The occulter (or planet) boundary integral $\mathcal{P}(\mathbf{G}_n)$ also starts with the intersection points. We define an angle $\phi$ to parametrize the bounds of the line integral. This angle, as noted in \citet{dholakia2022}, is defined similarly to an eccentric anomaly; the angle from the semimajor axis of the planet to the perpendicular projection of an intersection point onto the circle bounding the ellipse. Then, for the integrand, we start with the parametric formula for an ellipse:

\begin{align}
    x &= r_{eq} \cos(\phi) + x_o \\
    y &= r_{eq} b \sin(\phi) + y_o
\end{align}
We then plug this into the integrand for $\mathcal{P}(\mathbf{G}_n)$ in Equation~\ref{eq:pandq} to obtain:
\begin{align}
\mathcal{P}(\mathbf{G}_n) = \int_{\phi}^{2\pi + \phi}[\ G_{ny}(r_{eq} c_\phi + x_o , r_{eq} b s_\phi + y_o) b c_\phi] \ r_{eq} d\phi \\
- \int_{\phi}^{2\pi + \phi}[G_{nx}(r_{eq} c_\phi + x_o, r_{eq} b s_\phi + y_o)s_\phi]\ r_{eq} d\phi
\end{align}
where we write $\sin{(\phi)}$ as $s_\phi$ and $\cos{(\phi)}$ as $c_\phi$ for brevity.

\section{Implementation in eclipsoid} \label{sec:eclipsoid}

We implement the method in Section~\ref{sec:model} in the open-source Python package \href{https://github.com/shishirdholakia/eclipsoid}{\texttt{eclipsoid}}, which is written using the \textsc{Jax} framework \citep{jax}, which provides just-in-time compilation and automatic differentiation. We have designed the code as an extension to \href{https://github.com/exoplanet-dev/jaxoplanet}{\texttt{jaxoplanet}} \citep{jaxoplanet}, and inherit much of the user interface and some of the low level routines, such as the orbital solver, directly from \texttt{jaxoplanet}.

Users have two options to parametrize the orbit of an oblate occulter (as in \texttt{jaxoplanet}); a \texttt{TransitOrbit}, which uses observable quantities such as the transit duration and impact parameter. Alternatively, users can specify a \texttt{System}, which contains an \texttt{Central} object, which is assumed to be circular in projection, and a \texttt{Body}, along with Keplerian orbital parameters such as the inclination, argument of periastron, etc. The \texttt{TransitOrbit} is a useful parametrization for sampling for observed quantities instead of possibly unknown orbital parameters, e.g in the case of a single transit observation. Either of these orbits can be passed into a \texttt{limb\_dark\_oblate\_lightcurve} method, which additionally takes the limb darkening coefficient vector $u$, the projected oblateness parameter $f$ and the projected obliquity $\theta$, assumes a circular equivalent radius as in Equation~\ref{eq:r_circ}, and returns a function to compute the light curve at an array of times. The API has been designed to extend to the computation of light curves for arbitrary ellipsoids, including tidally distorted bodies, oblate but precessing bodies, and bodies with surface maps using a similar interface to the \texttt{starry} framework. This can be done similarly to above, by passing \texttt{Central} and \texttt{EllipsoidalBody} objects into an \texttt{EclipsoidSystem}. Users can also optionally pass in \texttt{Surface} objects representing the map of the bodies represented as a decomposition in spherical harmonics. This \texttt{EclipsoidSystem} object can then be passed into the method \texttt{eclipsoid\_light\_curve}, which will return a function to internally rotate the central body, the ellipsoidal body and their respective surface maps into the correct viewing orientation at each time passed in, perform the integrals described in Section~\ref{sec:model}, and output the light curve. 

Starting with the user-provided orbit quantities, we use the following method in our implementation of \texttt{eclipsoid}. We modify the implementation of \texttt{jax.numpy.roots} to allow differentiation and apply it to solve the quartic equation in Equation~\ref{eq:quarticform}. The resulting complex roots are then checked and discarded if the imaginary component exceeds $10^{-10}$, which we find balances numerical instabilities in the root finding with avoiding falsely detecting intersection points. We then sort the intersection points such that the integration is performed correctly around the closed region depicted in Figure~\ref{fig:integral_bounds}. This is done by checking if the angle $\frac{\phi_0 - \phi_1}{2}$ between the two intersection points lies inside the unit circle; if it does, the two intersection points are correctly sorted clockwise about the origin. If it does not, the intersection points are switched. We finally perform the numerical integration using an efficient implementation of Gaussian quadrature with a default 30 sample points, which is sufficient to reach a precision of better than $10^{-8}$. This integration is repeated for each term in Green's basis. We then take the vector dot product with the limb darkening coefficients, which have been transformed into coefficients in Green's basis. This entire procedure is efficiently vectorized over every given time point in the light curve using the \texttt{vectorize} utility in \texttt{jaxoplanet}, which internally uses \texttt{jax.vmap}. 

As a test of the low-level implementation, we test the numerical integral against a brute-force integration of each term in Green's basis. For a fiducial planet with $f=0.5$, $\theta=33^\circ$, $b_o=0.95$ and $r_{eq}=0.15$, we construct a $3999\times3999$ grid. For each point on the grid, we check if it is under the occulter and under the occulted body, in which case it is multiplied by the area element and summed together with the other points. The results of this simulation 
demonstrate that our implementation matches the brute force integral for all terms in Green's basis to within the precision of the grid. In addition, our implementation matches the code \texttt{squishyplanet} in the limit of a radially-symmetric stellar surface map to within the precision of the numerical integration. 

\begin{figure}[ht!]
    \script{partials.py}
    \begin{centering}
        \includegraphics[width=\linewidth]{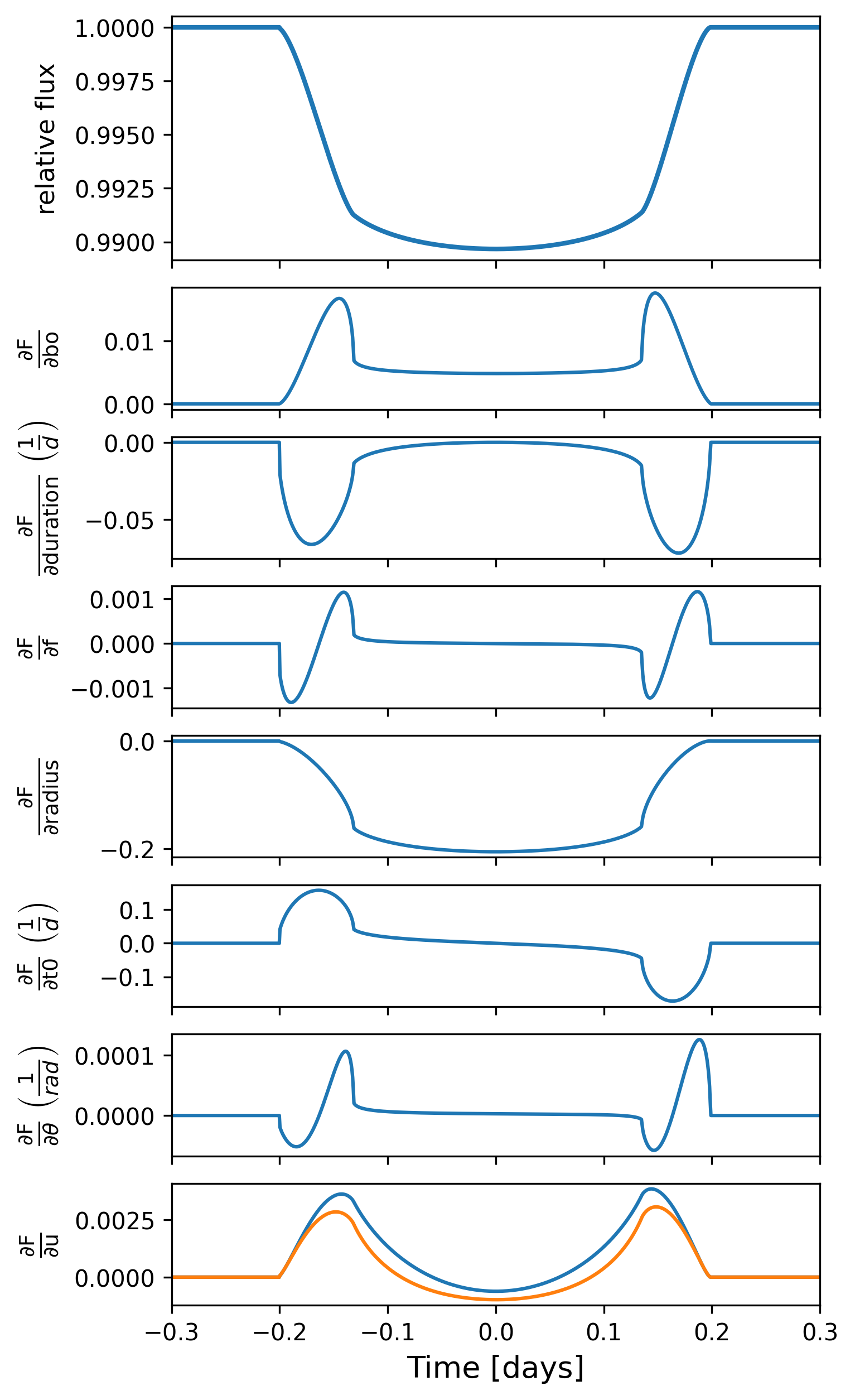}
        \caption{Model transit light curve showing partial derivatives with respect to each parameter in the \texttt{TransitOrbit} over time as computed by \textsc{Jax}. For the limb darkening derivatives, blue refers to $u1$ and orange refers to $u2$. \href{https://github.com/shashankdholakia/oblate-planets-paper/blob/main/src/scripts/partials.py}{\faGithub}}
        \label{fig:partials}
    \end{centering}
\end{figure}

We also tested the partial derivatives with respect to each parameter in the \texttt{TransitOrbit} to ensure that the model has stable first and second derivatives. Taking second-order gradients is important for several applications, including Fisher information and variational inference, which rely on the Hessian of the likelihood. In Figure~\ref{fig:partials}, we show partial derivatives with respect to each parameter in the \texttt{TransitOrbit} class computed for a fiducial planet with $f=0.1$, $\theta=35^{\circ}$, $b_o=0.7$, $u=[0.3, 0.2]$ and $r_{\mathrm{circ}}=0.1$. 

We benchmark the speed of our algorithm on a light curve with 1000 data points. Using \texttt{jax.jit} to compile away overheads in the function, we find that it takes roughly 7\,ms to compute the full light curve on an Mac M2 CPU. This corresponds to a speedup of about 2 orders of magnitude from existing implementations in \citet{carter2010a}, about 3 orders of magnitude in precision. 

\newpage 
\section{Prospects for Detection with JWST and other Instruments} \label{sec:jwstdetect}
In this section, we describe the use of our ellipsoidal model on simulated data from JWST to make inferences about the detectability of oblateness and obliquity. In particular, given the sample of known exoplanets, we wish to know which ones are the most sensitive to constraints on oblateness and obliquity with a single transit observation from JWST. For this kind of study, the white light photometric precision is of primary importance, and hence it is advantageous to have a lower spectral resolution as it avoids dispersing the light over too many detector pixels and contributing noise. The lowest resolution mode in JWST is NIRSpec PRISM \citep{nirspec}, but this mode saturates systems brighter than about 10$^\text{th}$ magnitude in $J$ band. The Near Infrared Slitless Spectrograph \citep[NIRISS;][]{niriss} on JWST provides a resolving power of $R \approx 1000$ with the GR700XD grism, and has been shown to be the most precise mode for other applications requiring broadband photometric precision in near infrared, such as eclipse mapping \citep{boone2024}. 

When designing an experiment to measure the oblateness and obliquity of a planet using the transit method, it is necessary to know whether the planet's transit light curve will provide information on these parameters. The primary challenge here is a spherical planet of equivalent area closely matches the transit of an ellipsoidal exoplanet. Hence, many works determine the detectability of oblateness in terms of the difference in signal between the ellipsoidal planet and an equal-area or best-fitting circular planet \citep{seager2002a, barnes2003, carter2010a, zhu2014}. However, this neglects the influence of other uncertain parameters of the model--such as the limb darkening coefficients or the impact parameter--on the oblateness and obliquity. To propagate all these sources of uncertainty onto the oblateness and obliquity, we perform a injection-and-recovery exercise, in which we simulate an oblate exoplanet transit with a fiducial set of parameters and attempt to recover the posterior distributions over the parameters of interest. 

We use \texttt{JexoSim 2.0} \citep{sarkar2021} to simulate JWST NIRISS SOSS observations of every transiting exoplanet on the Exoplanet Archive as of July 2024 with $P_\text{orb}>30$\,d, host $J_\text{mag}<13.0$ and $R_J>0.5$ with the GR700XD grism. For each exoplanet, we use the out of transit mode to simulate observations of a set duration with all noise sources, coming up with a total noise estimate. We then construct a light curve from a transit signal with an oblateness factor $f=0.1$. We use broad, uninformative priors on all parameters except the transit duration and limb darkening, which we set to log-normal and normal distributions respectively. For limb-darkening coefficients, we obtain the priors from following the approach used in \texttt{JexoSim}, using \texttt{ExoTethys} \citep{morello2020}. We obtain distributions of stellar temperatures and compute distributions of $\log g$ from stellar mass and radius from the latest values provided in the Exoplanet Archive (as of July 2024). We then propagate these distributions to the limb-darkening coefficients and transform them according to \citet{kipping2013} to get our normal LDC priors in sample space. We then use the NUTS sampler, which is an implementation of Hamiltonian Monte Carlo  in NumPyro \citep{phan2019}, to recover the posterior distribution for that particular simulation, assuming white noise equivalent to the noise level found in the \texttt{JexoSim} simulation. This process is performed at each of several obliquities $\theta \in \{0^\circ, 20^\circ, 40^\circ, 60^\circ, 80^\circ\}$, and for each exoplanet in the sample. We do not include values of $\theta$ below zero as these can be mapped to the former values by symmetry. 

In our simulations, we initially attempted to use the full transit simulations in \texttt{JexoSim}, which take into account correlated noise sources. However, we found that the computational cost was much higher for these simulations, the results were not significantly different, and our approach allowed us to minimize the effect of biases from a particular noise realization. We tested each chain for convergence, and re-ran any simulations that had fewer than 100 effective samples per chain or had \citet{gelman92} statistic $\hat{r}>1.1$.

\begin{figure}[ht!]
    \script{mcmc_vs_fisher_info.py}
    \begin{centering}
        \includegraphics[width=\linewidth]{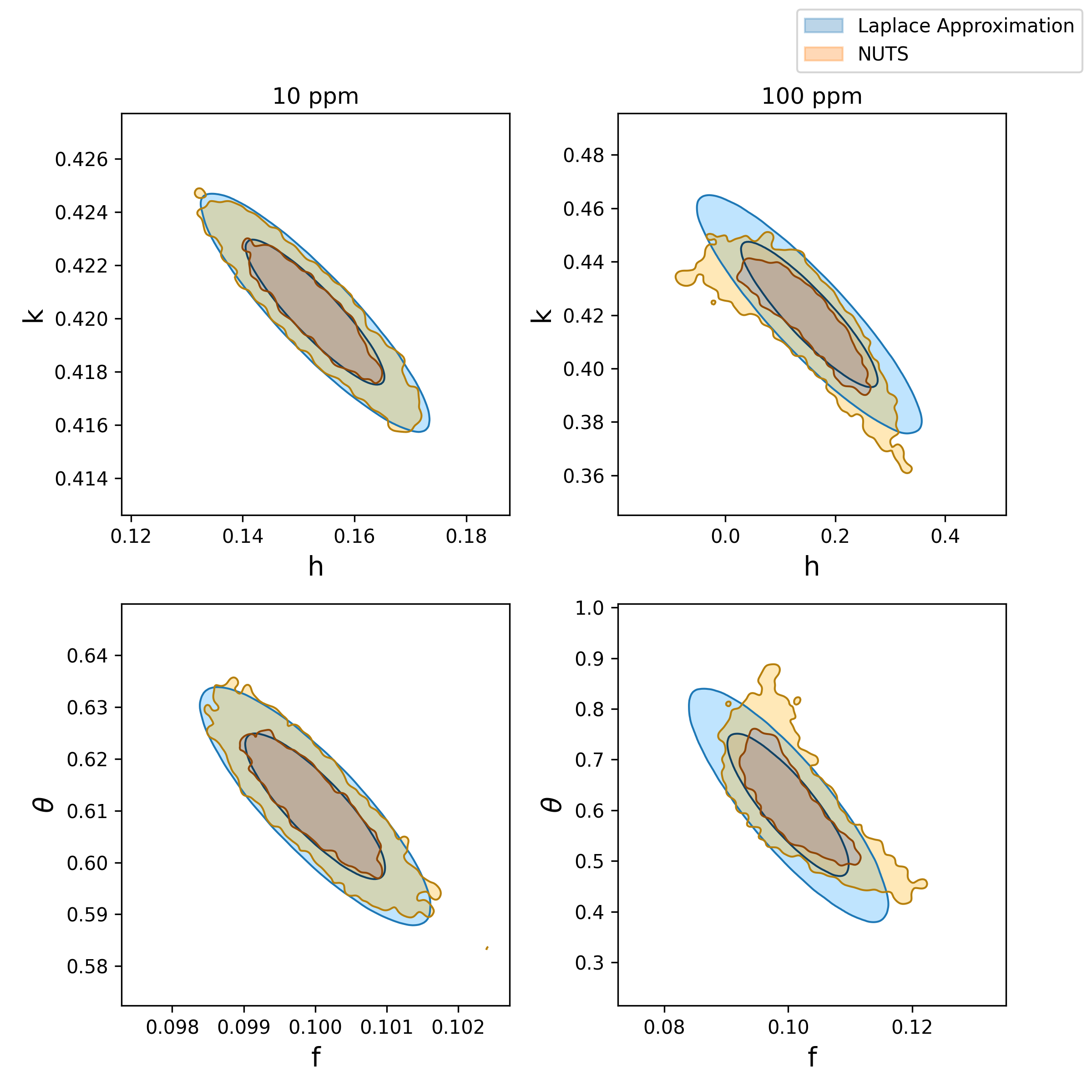}
        \caption{Comparison of a posterior obtained from MCMC sampling of a simulated transit observation (orange contour) with that obtained analytically from the Laplace approximation (blue contour) for two fiducial injected noise values, 10\,ppm (left) and 100\,ppm (right). \href{https://github.com/shashankdholakia/oblate-planets-paper/blob/main/src/scripts/noise_mcmc_fisher_plotter.py}{\faGithub}}
        \label{fig:fishercorner}
    \end{centering}
\end{figure}

When attempting to fit for the oblateness and obliquity, it is helpful to understand the shape of the posterior on the transit parameters. We also attempted to recover posterior distributions keeping the transit parameters constant but at two different noise values and two parametrizations of the oblateness and obliquity parameters. We find that in the limit of very high SNR ~10\,ppm, the posterior is well approximated by a Gaussian centered around the point of maximum likelihood, as described by the Laplace approximation. As the noise is increased to 100\,ppm or greater, the posterior becomes increasingly non-Gaussian, as seen in Figure~\ref{fig:fishercorner}.

In addition, the $f$and $\theta$ parametrization presents problems when performing inference. Firstly, there are significant covariances between the two parameters and strong non-Gaussian degeneracies. Secondly, at $f=0$, the parameter $\theta$ is undefined and so oblateness values of zero encounter a singularity close to 0. These issues are resolved by using the parameterization
\begin{equation}
    \begin{aligned}
        h &= 2 \sqrt{f/2} \cos{2 \theta} \\
        k &= 2 \sqrt{f/2} \sin{2 \theta}
    \end{aligned} \\
    \quad f \in [0, 0.5],\ \theta \in \left[-\frac{\pi}{2}, \frac{\pi}{2}\right]
\end{equation}

When sampling in the parameters $h$ and $k$, an additional benefit is that when these parameters are uniformly sampled between 0 and 1, the corresponding $f$ and $\theta$ are uniform on a half-disk, naturally bounding the values of $f$ and $\theta$ to physically realistic and non-degenerate values. 

We are interested primarily in the two parameters, $f$and $\theta$, or the oblateness and obliquity respectively, where all the other transit parameters are marginalized over. As $\theta$ is an angular parameter, some samples are wrapped over the range, so it does not admit a simple summary statistic like standard deviation. In addition, any summary statistic on either one of those parameters only presents half the picture, as there are cases where one parameter is strongly constrained at the cost of the other. In such situations where one is interested in the total constraint on two or more parameters, it is common to use some function (such as the determinant) on the eigenvalues of the Fisher information matrix \citep{Coe2009}, the inverse of which is the covariance matrix about the maximum likelihood value. This can informally be thought of as measuring the area of the elliptical contour formed by two covariant parameters. 

However, in our case, as demonstrated above, the parameters are highly non-Gaussian, and using the Fisher information matrix would miss the highly curved posterior distribution around $f$and $\theta$. There are two ways to consider the total constraint from such a highly curved distribution around $f$and $\theta$. The first approach is pessimistic: due to the highly degenerate values of $f$and $\theta$ that are allowed by the fit, it is not possible to come to a unique constraint on either parameter, and hence the area should large enough to encapsulate the region of space allowed by the fit on either parameter. For this approach, we find the convex hull containing 95\% of the samples in $f$ and $\theta$ space to summarize the information. The other way to think about it is in terms of ruled-out parameter space. A highly narrow, curved and covariant distribution, although failing to provide unique values of either parameter, nevertheless rules out most regions in parameter space, so is highly constraining. To summarize the information using this approach, we use the area of the 95\% contour of a kernel density estimate. 

\begin{figure*}[ht!]
    \begin{centering}
        \includegraphics[width=\linewidth]{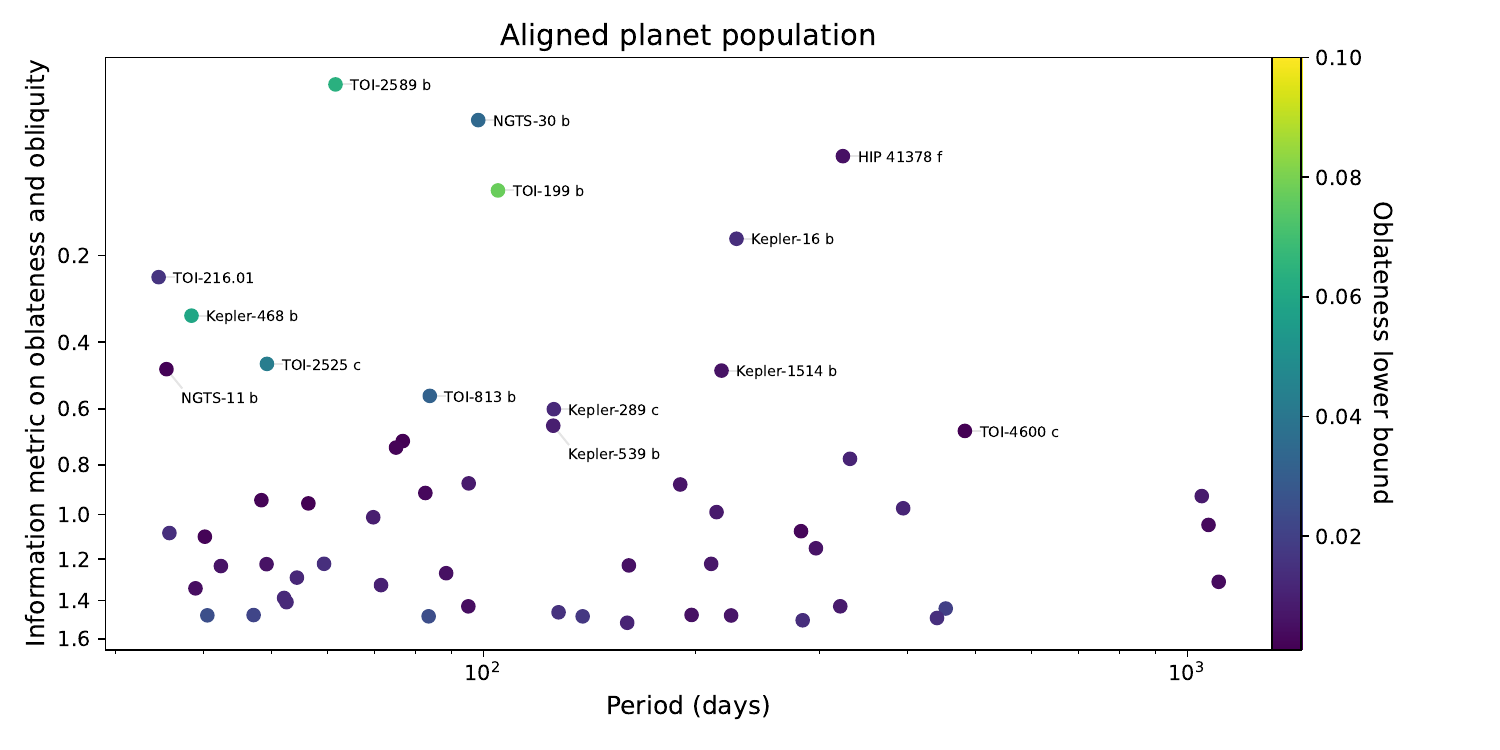}
        \includegraphics[width=\linewidth]{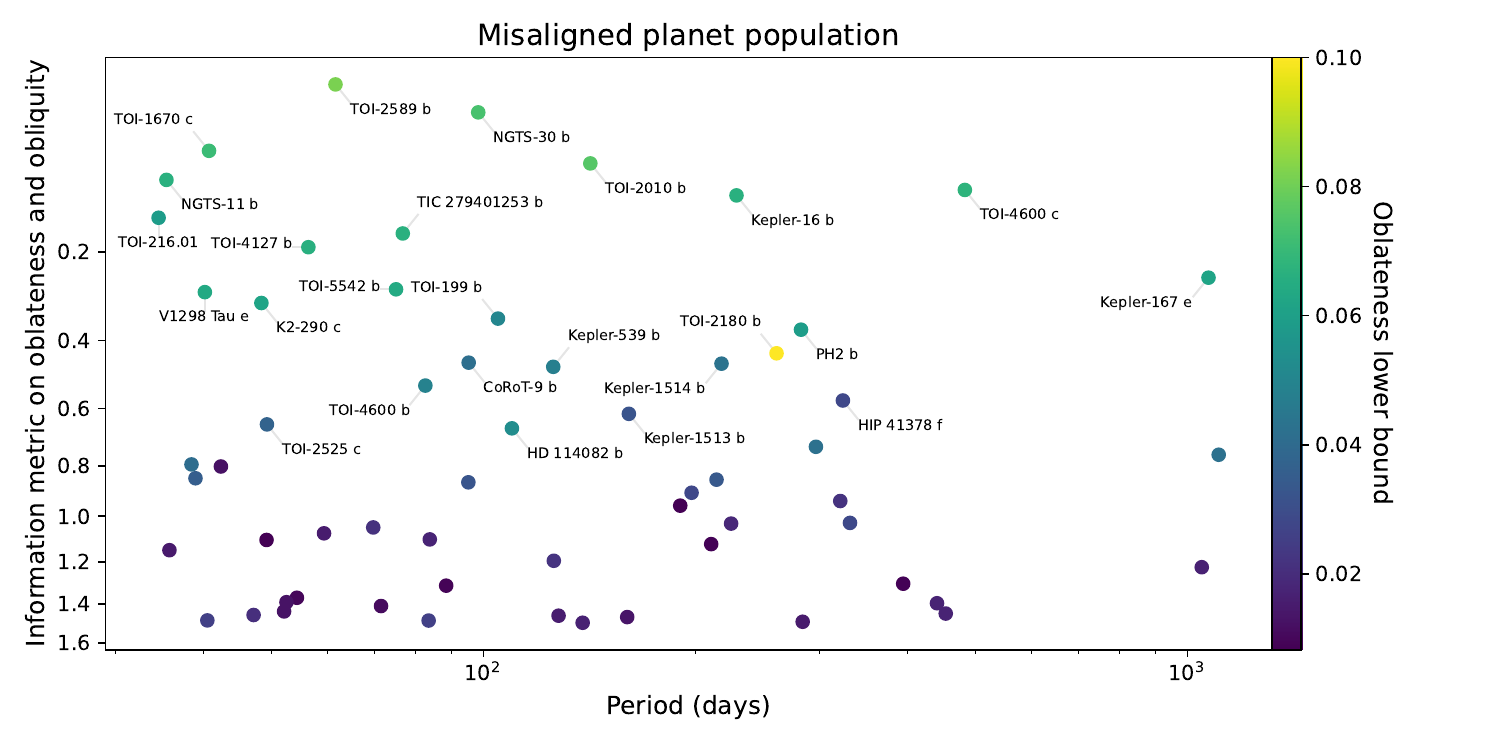}
        \caption{Information content on oblateness and obliquity across a simulated planet population where all planets have a Saturn-like oblateness $f=0.1$. We define the information metric as the 95\% convex hull area on $f$ and $\theta$ vs. period for an aligned planet population (top) vs misaligned (bottom). Colors represent the 5th percentile on $f$ in each population. Aligned population is the $\theta=0^{\circ}$ case in the simulation; misaligned case is the average of the results from $\theta=20^{\circ}$, $40^{\circ}$, $60^{\circ}$ and $80^{\circ}$. Targets with stronger constraints on oblateness and obliquity are higher on the y-axis, and targets with lighter colors represent higher detectability of rotational deformation. \href{https://github.com/shashankdholakia/oblate-planets-paper/blob/main/src/scripts/jwst_sim/param_space_plotter.py}{\faGithub}}
        \label{fig:parameter_space}
    \end{centering}
\end{figure*}

We show the full planet population in Figure~\ref{fig:parameter_space}. We bin the simulations into two categories under the assumption that the planet population is mostly aligned or that the planet population is isotropically misaligned. For the aligned case, we use only the $\theta=0^{\circ}$, whereas for the misaligned case, we average the results of the simulation for $\theta\in\{20^{\circ}, 40^{\circ}, 60^{\circ}, 80^{\circ}\}$. These results are meant to illustrate the expected information content on oblateness and obliquity from a single JWST observations of any of these planet populations for both aligned and misaligned geometries.

\begin{figure}[ht!]
    \begin{centering}
        \includegraphics[width=1.05\linewidth]{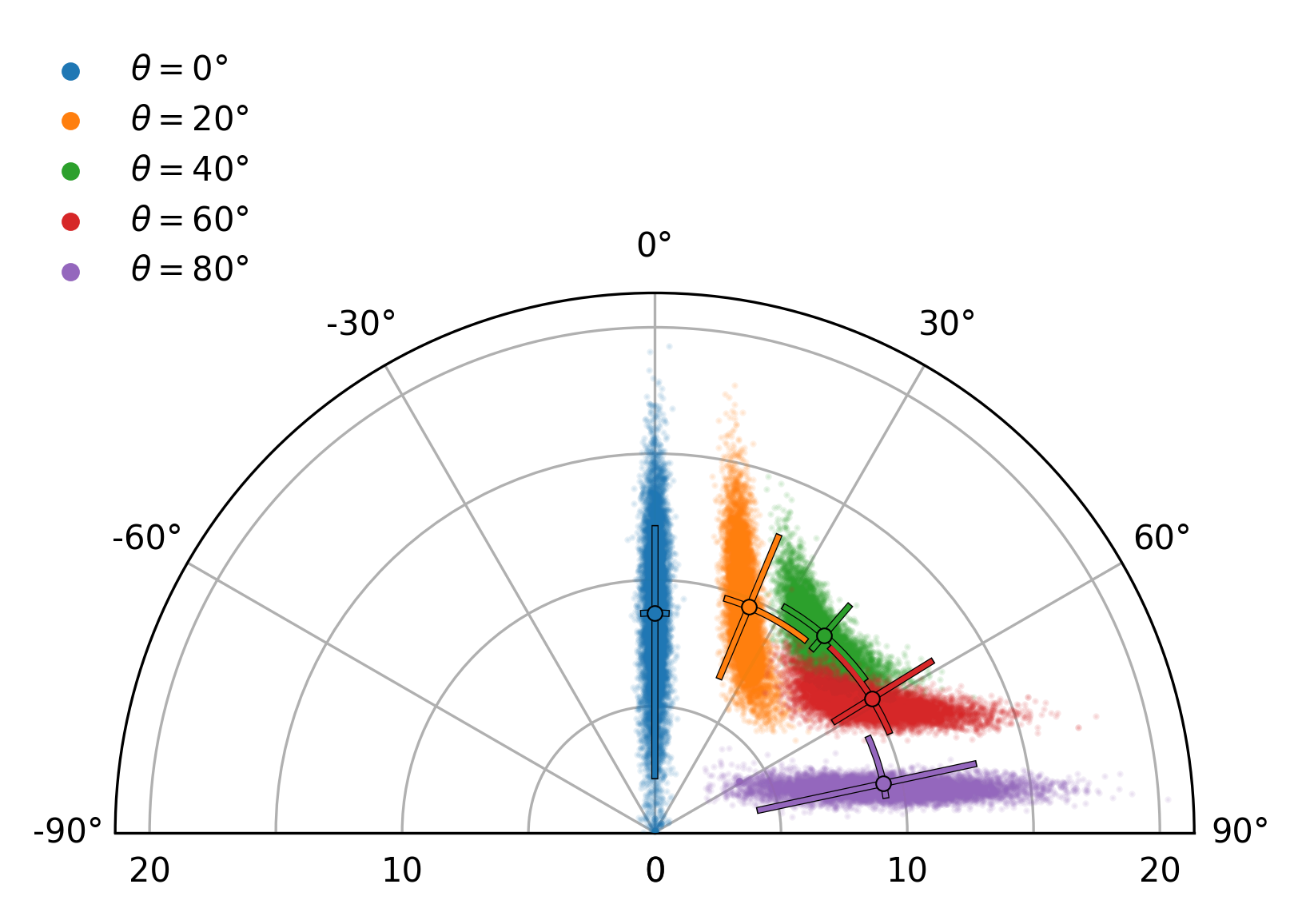}
        \caption{Polar plot of simulated posterior samples of NGTS-30\,b, with $f$ as the radial coordinate and $\theta$ as the angular coordinate. Black filled circles mark the mean recovered value (using the circular mean for the angular parameter $\theta$), error bars represent the 5th and 95th percentiles on the relevant quantities, and colors represent each injected obliquity value. \href{https://github.com/shashankdholakia/oblate-planets-paper/blob/main/src/scripts/jwst_sim/plot_polar_contours.py}{\faGithub}}
        \label{fig:ngts30polarplot}
    \end{centering}
\end{figure}

\subsection{Effect of Impact Parameter}
\begin{figure}[ht!]
    \begin{centering}
        \includegraphics[width=1.05\linewidth]{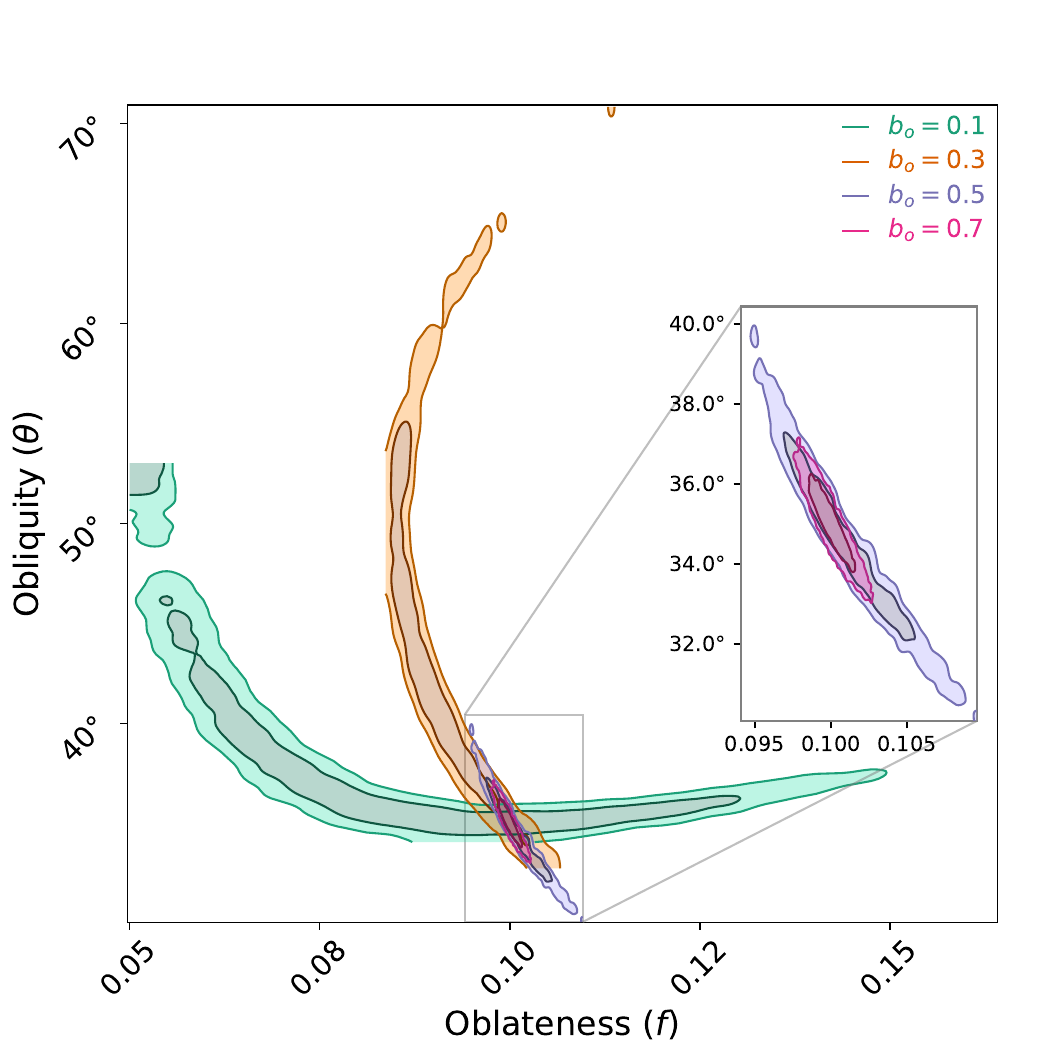}
        \caption{Comparison of MCMC posterior contours on projected oblateness and obliquity for a simulated transit with four different values of impact parameter but otherwise identical parameters. Zoomed inset shows $b_o=0.5, 0.7$, where the degeneracy contour is much smaller and less curved. High impact parameters yield significantly lower degeneracies in oblate planet fits. \href{https://github.com/shashankdholakia/oblate-planets-paper/blob/main/src/scripts/impact_parameter_comparison.py}{\faGithub}}
        \label{fig:impactparamcomparison}
    \end{centering}
\end{figure}

In our simulations on the full sample of parameters, we found that there exists a significant correlation between the impact parameter of a transit and the achievable precision on oblateness and obliquity. Planets with high impact parameters in the range of $0.7<b_o<0.8$ have the most uniquely constrained oblateness and obliquity, so much so that the impact parameter can drown out other factors such as the planet radius and host magnitude. This effect is particularly notable as it allows unique constraints even when the planet is nearly or fully \emph{aligned}, whereas in most other cases an aligned, oblate planet is degenerate with a spherical planet with a different impact parameter, and a range of different oblateness values in between. An example of this is  NGTS-30\,b or TOI-2589\,b, both of which have high impact parameters and also have the best constraints on oblateness and obliquity in our simulations. For a fiducial planet with radius=0.14 and a period of ~7 days, we inject an oblate transit with 1000 points with four different impact parameters $b_o=0.1, 0.3, 0.5, 0.7$ but otherwise identical parameters ($f=0.1$, $\theta=35^\circ$) and a point-to-point scatter of 10\,ppm. We then recover the posterior distributions on $f$ and $\theta$ using Hamiltonian Monte Carlo with \texttt{NumPyro}. We show the resulting contours in Figure~\ref{fig:impactparamcomparison}. The contours on $f$ and $\theta$ at low impact parameters are significantly more degenerate and non-Gaussian, even for this simulated example with high signal-to-noise. 

Both for a given set of parameters, and across the exoplanet population, planets with higher impact parameters tend to be more sensitive for studies of axial tilt and rotational deformation. We discuss the implications for this in further in Section~\ref{sec:disc}.

\newpage

\section{Application to JWST NIRSpec transit of WASP-107~b}
\label{sec:lctest}
\begin{figure}[ht!]
    \begin{centering}
        \includegraphics[width=\linewidth]{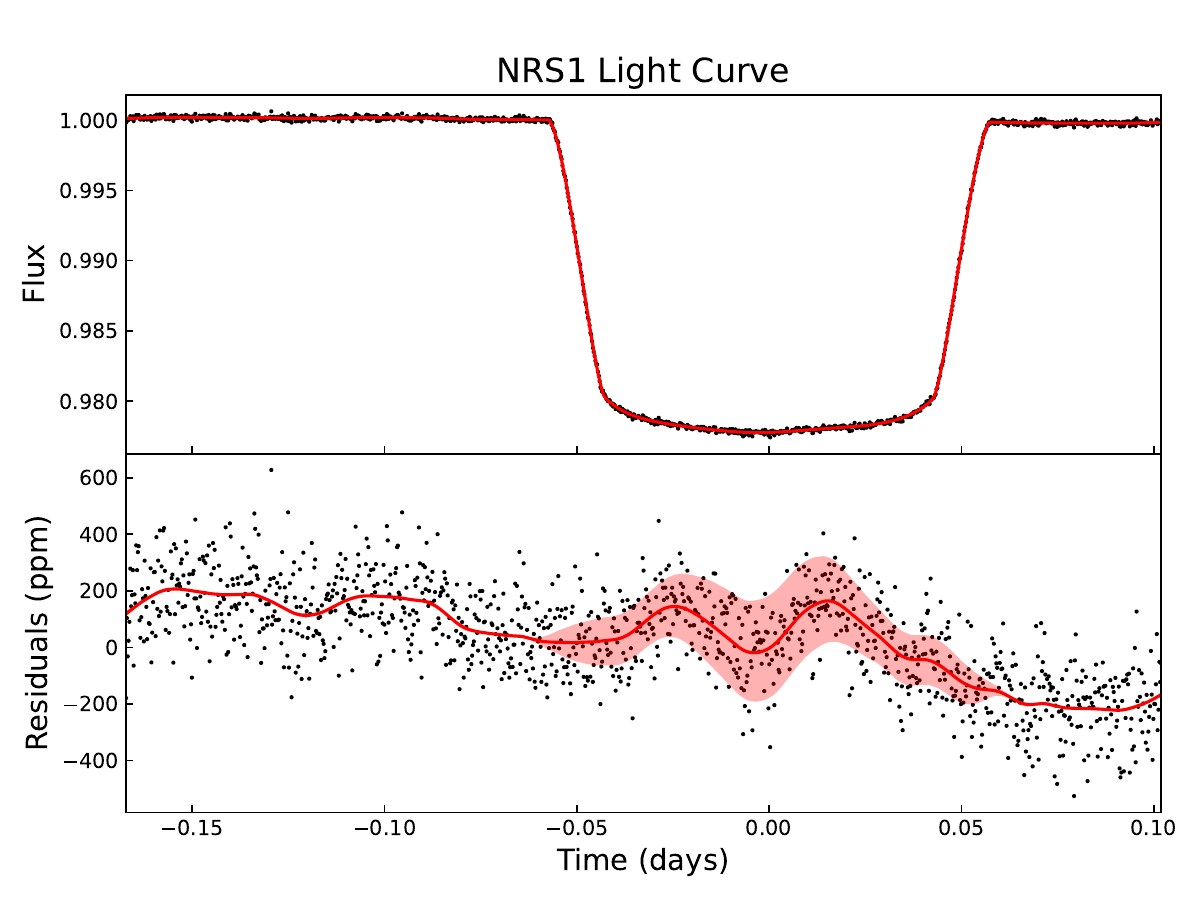}
        \includegraphics[width=\linewidth]{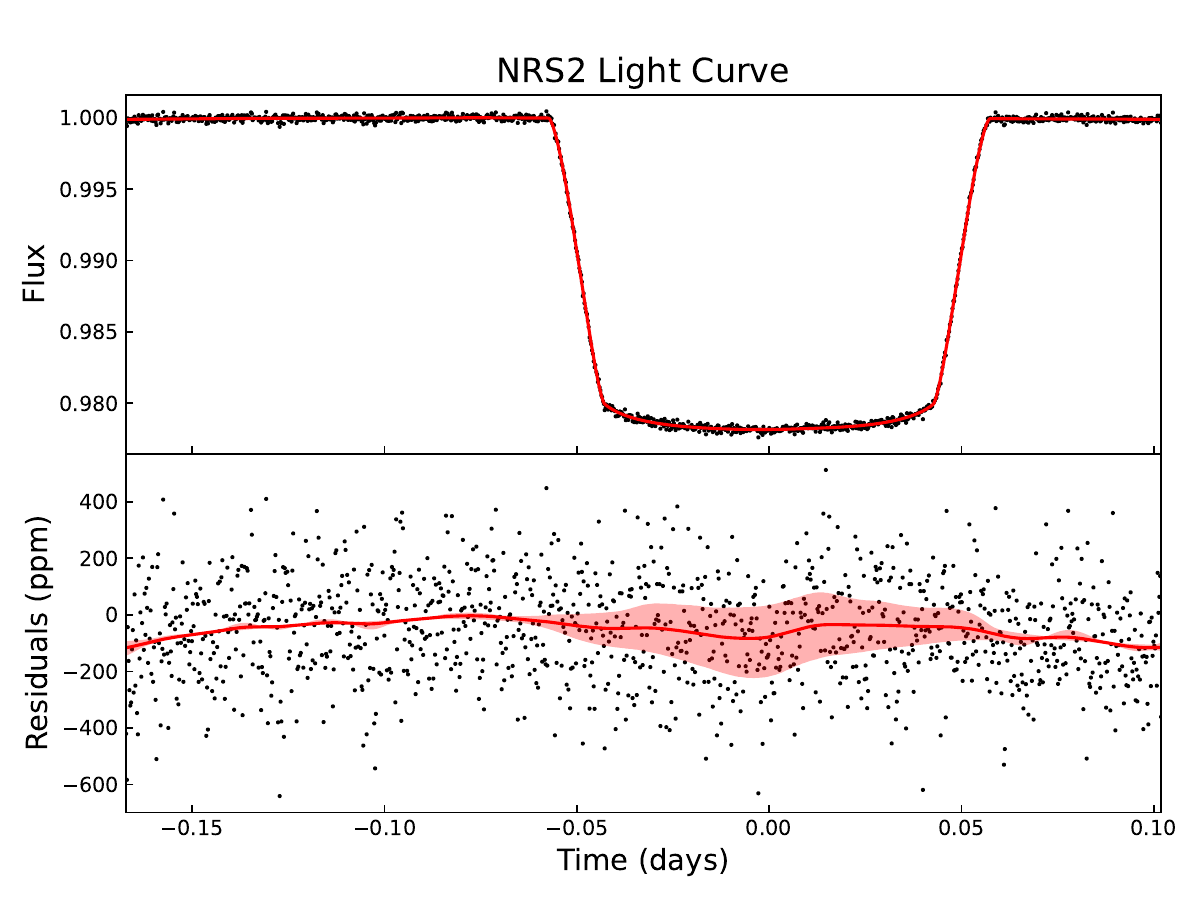}

        \caption{JWST light curve of WASP-107~b. The top and bottom figures are the NRS1 and NRS2 detectors respectively. The top subplots show the full transit with a red line for the mean transit model from the MCMC fit. The bottom subplots show the residuals of the oblate transit fit as a scatter plot. The GP mean model is overplotted as a red line, with the red shaded area representing the $5^\text{th}$ and $95^\text{th}$ percentiles of the GP. \href{https://github.com/shashankdholakia/oblate-planets-paper/blob/main/src/scripts/wasp107~b/WASP107_NUTS_GP.ipynb}{\faGithub}}
        \label{fig:wasp107lc}
    \end{centering}
\end{figure}

In order to test the model on a real JWST transit observation, we fit the NIRSpec G395H observations of WASP-107~b presented in \citet{Sing2024}. WASP-107~b is a low-density, Jupiter-sized planet with a 5.7 day orbital period around a K type star. We fit an oblate model with \texttt{eclipsoid} jointly to the NRS1 (2.7-3.71~$\mu$m) and NRS2 (3.83-5.16~$\mu$m) broadband light curves of WASP-107~b. We use broad, uninformative priors on the limb darkening from the physically motivated reparametrization of limb darkening in \citep{kipping2013}. We adopt broad or uninformative priors on all other parameters except period, which we keep fixed in the single transit observation. We also fit a Gaussian Process (GP) using a quasiseparable Matern-3/2 kernel \citep{dfm2017, aigrain2023} along with the oblate transit model in order to detrend the correlated noise particularly present in the NRS1 channel, and add in quadrature a separate jitter term for each light curve to account for unnaccounted error sources. We use Hamiltonian Monte Carlo \citep[HMC;][]{Betancourt2017} with \texttt{NumPyro} to sample 4 chains each with 4000 samples, discarding the first 2000 as burn-in. The results of the fit are presented in Figure \ref{fig:wasp107lc}. 

\begin{figure}[ht!]
    \begin{centering}
        \includegraphics[width=\linewidth]{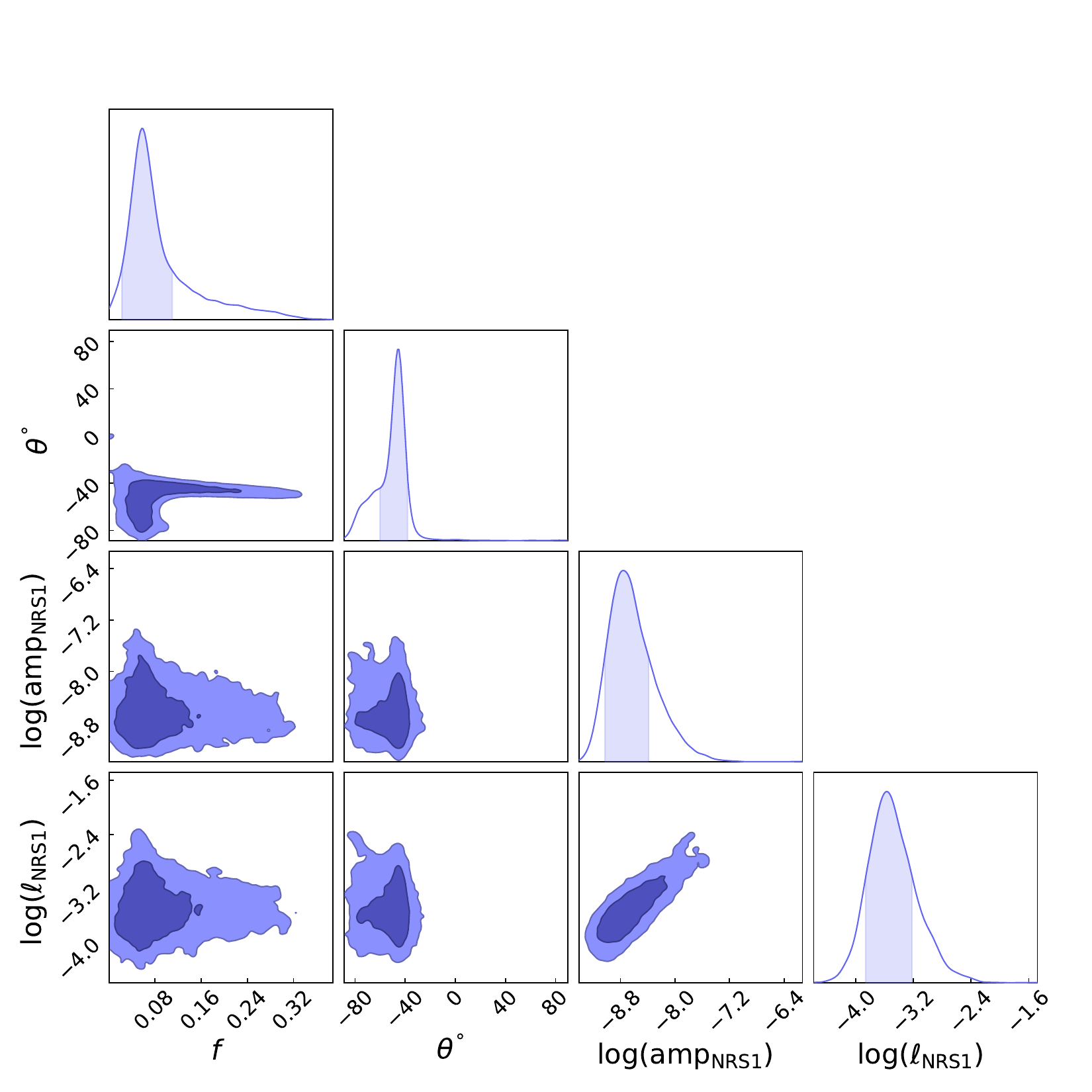}
        \caption{Reduced posterior contours on oblateness and obliquity for an oblate model plus GP fit to the JWST NIRSpec transit of WASP-107~b. Included are the oblateness and obliquity $f$ and $\theta$, as well as the GP hyperparameters for NRS1, where the correlated noise is most significant. No significant correlation is seen between $f$ and $\theta$ and the GP hyperparameters, implying that the correlated noise strongly bias the recovered oblateness and obliquity. An extended version of this corner plot, including all fit parameters, is available online.  \href{https://github.com/shashankdholakia/oblate-planets-paper/blob/main/src/scripts/wasp107~b/WASP107_NUTS_GP.ipynb}{\faGithub}}
        \label{fig:corner_oblate_gp}
    \end{centering}
\end{figure}

We obtain an oblateness factor of $f=0.07^{+0.08}_{-0.03}$, consistent with zero oblateness. This is expected for a planet with a 5.7 day orbital period; the expected oblateness factor if WASP-107~b is tidally locked to its host star would be $f \approx 0.002$. Taking the 5th percentile of the oblateness posterior as an upper bound on the projected oblateness ($f<0.23$), WASP-107~b's rotation period must be $P_{\mathrm{rot}}>13$\,h assuming that the planet is equator-on with respect to our line of sight. The GP hyperparameters are not strongly covariant with the oblateness or obliquity, suggesting that the GP does not strongly affect the recovered distributions on $f$ and $\theta$; we show the posterior contours along with the GP hyperparameters for NRS1, where the correlated noise is more significant, in Figure~\ref{fig:corner_oblate_gp}, while the full corner plot of all parameters is available online.

\section{Discussion and Future Work} 
\label{sec:disc}
We introduce a fast, differentiable model for computing light curves of ellipsoidal planets orbiting stars with arbitrary surface maps. We implement this model in the open-source \textsc{Jax}-based package \texttt{eclipsoid}, with the goal of providing the community a tool to model rotational deformation in planetary transit observations. Using the \texttt{eclipsoid} model, we inject an oblate transit into simulated JWST NIRISS/SOSS observations for each of the entire known sample of bright, long-period transiting planets. We recover the posterior distributions for the oblateness and obliquity, and demonstrate that JWST is sufficiently sensitive to detect these effects for the first time for a growing sample of planets. We also show the best targets for JWST characterization of oblateness and obliquity in data that may be additionally used for atmospheric characterization. 

The implementation of the model in a differentiable framework extending the \texttt{jaxoplanet} package has several implications for characterizing transiting exoplanets beyond just specific studies of oblateness and obliquity. The model is, for the first time, fast and general enough to be used on transit fits while marginalizing over oblateness and the stellar surface map as unknown parameters. The ability to take gradients with respect to the parameters of the model is important for cases where there are over a dozen parameters, in which cases non gradient-based techniques such as affine invariant sampling \citep{Goodman2010} or nested sampling \citep{Skilling2004,Buchner2023} can struggle or fail to converge \citep{Huijser2015}. 

One envisioned use for this model could be, for instance, in atmospheric transmission spectroscopy, where impact parameter and other transit parameters are fit jointly with the effective radius of the planet in each wavelength bin. For long period planets, it may be important to marginalize over unknown oblateness and obliquity in transmission spectroscopy data. More work is needed to determine whether these effects could bias transmission spectra or cause underestimated errors. 

Another use case for our model is in marginalizing over the stellar surface map when attempting to determine the oblateness and obliquity of an exoplanet. The recent study by \citet{lammers2024} on Kepler-51d, in which case a large starspot occultation is found in the transit, is a good example. With the increased precision of JWST, starspot crossings and starspot modulation are significant effects and have already shown concern for biasing transmission spectra \citep[the Transit Light Source Effect;][]{Rackham2018}; it is therefore relevant to be able to model and marginalize over such spots while fitting for rotational deformation, especially during ingress and egress where most of the information on oblateness and obliquity lie. 

\subsection{Parameter Estimation or Model Comparison?}

Previous studies on planetary oblateness and obliquity generally use one of two frameworks for studying light curves that may bear signatures of planetary deformation: model comparison or parameter estimation. In the model comparison framework, a technique that estimates the Bayesian evidence integral such as nested sampling is used. The evidence computed from the oblate planet model is then compared to the evidence computed for the circular planet model to decide whether the detection is significant and which model is more parsimonious with the data. The other framework that is commonly used is parameter estimation, where only an oblate model is used, and the circular planet case is seen as a subcase of the oblate planet model for which $f=0$.

Which framework is the correct one for the problem of studying rotational deformation in exoplanets? It is inevitable that model comparison will find uses in studies of planets with JWST and other realistic datasets, as it is important to show that the effect of oblateness is large enough to justify using the extra parameters. However, we emphasize that there is a fundamental difference between the oblateness of an exoplanet and other similar cases where model comparison would find use, such as in assessing evidence for the presence of an additional planet in an exoplanet system, or in estimating the flatness of spacetime in cosmology. In the case of an additional planet, the model without an additional planet \emph{could} be seen as a subcase of the two planet model with the mass of the second planet set to 0. In the case of estimating the geometry of spacetime, the flat universe is a subcase of $\Lambda$CDM in which the total matter density is set to 1. 

In contrast to these example though, where the two models represent different versions of reality and it is not known \textit{a priori} which model is correct, in the case of oblateness, we \emph{know from physics} that all rotating bodies must be rotationally deformed at some level. It is then just a matter of estimating the extent of the deformation. This implies that parameter estimation is the correct framework for oblateness studies in general. This is especially true for long-period transiting planets, which as of yet have no strong constraints on rotation.

Nevertheless, if transiting exoplanets are not significantly oblate, which could be the case for planets which have been tidally sychronized, parameter estimation may lead to certain biases in the estimated oblateness due to the hard physical boundary at $f=0$. Such an effect has been noted for eccentricity estimation for binaries in circular orbits \citep{lucy1971}, an analogous case to oblateness estimation in multiple ways. The solution to such biases in the case of eccentricity is to bake in our knowledge of the physics of tidal dissipation as an informative prior \citep{lucy2013}. We propose that a similar prior may be warranted in the case of oblate planets, in which the tidal synchronization timescale for a given planet is propagated into an informative prior on the planet's oblateness. 

It bears noting that much of the interest in estimating the oblateness of planetary systems lies in probing physical effects that are not yet well understood, such as the effects of neighboring planets on tidal spindown \citep{millholland2019}. It may therefore still be instructive to assume little or no information on the oblateness (in the case of parameter estimation) and assert that a spherical planet model is indeed not more parsimonious with data (in the case of model comparison), at least until detections of oblateness become more common.

\subsection{Future Work}
The equations derived here, as in \texttt{eclipsoid}, include both oblate and prolate planets. While we have focused in this paper on inferring oblateness and therefore rotation, the same methods would be valuable in directly constraining the tidal deformation of close-orbiting planets. Because the projected value of prolateness for a tidally-locked exoplanet is very small in transit, the effect is comparatively much harder to detect than oblateness from a transit observation alone. With a full phase curve, the information during ingress and egress is combined with the projected area of the planet changing as a function of phase, which greatly increases the significance of a detection \citep{akinsanmi2024}. This makes it strongly advantageous to perform a full joint model over a phase curve, including the transit, eclipse, and out-of-transit variability due to the gravity-darkening effect on the star, the planet's changing projected area, and the planet's surface map. With the framework presented in this paper, it is possible with a gradient-based sampler like HMC to obtain posteriors over all of these parameters while marginalizing analytically over the star's and planet's surface map. We leave the details of this technique and application to JWST phase curves to a future work.

Similarly, any studies of rotational deformation have significant overlap with studies of circumplanetary rings; both effects are expected to be observed for giant transiting planets at longer orbital periods due to weak stellar tides. Where rings can be approximated as opaque and solid, such as for large, face-on rings, \texttt{eclipsoid} can also constrain their existence and geometry using our geometric model. The technique used in eclipsoid can also be used to model true ring geometries; by adding several elliptical annuli each with a given opacity, one would need to sort the resulting intersection points to make a closed region to integrate over, similar to the PyPplusS code \citep{rein2023}.

Lastly, this work, along with \citet{dholakia2022}, provide methods for analytically modelling occultations of oblate bodies where either the occultor or the occulted body are oblate, but not yet for the case where \emph{both} bodies are oblate. This is easily achievable through an affine transformation of the coordinate system such that one of the bodies is `popped' into a circular shape in projection, but the applications of this, at least in the field of exoplanets, are somewhat limited. This could, however, be a useful model for eclipsing binary stars where both bodies are tidally or rotationally deformed. A future work will extend \texttt{eclipsoid} to be able to model both the central and occulter bodies as oblate.

\subsection{Selecting Targets for JWST Observation}
In our simulations of the planet population (Section \ref{sec:jwstdetect}), we identify targets that would yield precise estimates of obliquity and oblateness in JWST NIRISS SOSS observations. We do not attempt to put any limits on detectability; any attempt to prove or disprove a detection in real data would likely require modeling JWST red noise in the light curves and using model comparison between oblate and spherical transit models in recovery. Our simulations provide a relative metric for the information as well as the expected precision of measurements of rotational deformation in JWST observations. 

In addition to identifying targets which would result in precise constraints on oblateness and obliquity, good targets for future oblateness/obliquity observations should be targeted at systems that are expected to be significantly rotationally deformed. Longer period or younger planets are less likely to have been tidally spun down and/or synchronized by their host star. Lower density planets may deform under lower rotational velocity, making oblateness and obliquity more observable for a given planetary spin rate. Finally, future studies of oblateness and obliquity should test theories of planetary formation, evolution, and demographics. 

Observations that can identify planetary oblateness and obliquity, such as NIRISS SOSS observations we have simulated, can also be used for other science goals. Target for oblateness studies may be more valuable if atmospheric characterization would also yield significant scientific value and be achievable  \citep[e.g. planets that have a high Transmission Spectroscopy Metric as defined in ][]{kempton2018}.

For the planet population simulated with orbit-aligned geometry, TOI-2598 b, NGTS-30~b, HIP 41378~f, TOI-199~b \citep{hobson2023}, and Kepler-16~b have the strongest constraints on oblateness and obliquity. For the population simulated with misaligned spin axes, TOI-2598~b, NGTS-30~b, TOI-1679~c, TOI-2010~b, and TOI-4600~c have the most precise constraints of oblateness and obliquity.

Of note, NGTS-30~b stands out as having a constraint on age based on a high $v\sin{i}$ measurement \citep{battley2024} as $<1\pm0.4\,$Gyr. Combined with our simulations showing that it has one of the strongest constraints on oblateness and obliquity regardless of its alignment, this makes NGTS-30~b one of the best targets for future follow up with JWST. The relatively young age and long period of this planet indicate a long tidal synchronization timescale \citep{rauscher2023}, and that studying the transit of this system would yield constraints on the primordial spin rate of NGTS-30~b, well before tidal effects have had a chance to synchronize its spin. 
HIP 41378 f is also a planet of note \citep{vanderburg2016}: as a super-puff planet, its low density is enigmatic and could yield detectable rotational deformation even for low planetary spin rates. Its apparent low density has also been attributed to opaque circumplanetary rings which may also be detectable \citep{akinsanmi2020,alam2022,belkovski2022,saillenfest2023,lu2024}. Opaque, face-on rings like in the long-period transiting system J1407 \citep{Mamajek2012,Kenworthy2015} appear elliptical in projection, which could be modeled in the same way as oblateness; gapped or partially-transmissive rings can be represented as linear combinations of these transit models. From the light curve of their sharp ring edges crossing spherical-harmonic spots it will be possible to simultaneously constrain the ring structure \textit{and} maps of the stellar surface.

\newpage
\section*{Acknowledgments}

We would like to acknowledge valuable discussions and input from Hugh McDougall, Abb\'{e} Whitford, Alexander Venner, Ellen Price, Andrew Vanderburg, Matthew Kenworthy, Wei Zhu, Ben Cassese, Chelsea Huang, Simon Murphy, and Duncan Wright. We also thank the authors of \citet{Sing2024} for making their reduced broadband light curves accessible to the community, making the analysis in Section~\ref{sec:lctest} possible.

Benjamin Pope and Shashank Dholakia were funded by the Australian Government through the Australian Research Council DECRA fellowship DE210101639.

We acknowledge and pay respect to the traditional owners of the land on which the University of Queensland and University of Southern Queensland are situated, upon whose unceded, sovereign, ancestral lands we work. We pay respects to their Ancestors and descendants, who continue cultural and spiritual connections to Country. 

This research has made use of the NASA Exoplanet Archive, which is operated by the California Institute of Technology, under contract with the National Aeronautics and Space Administration under the Exoplanet Exploration Program. 

\software{
Astropy \citep{astropy:2022}, \textsc{NumPy} \citep{numpy}; Matplotlib \citep{matplotlib}; \textsc{Jax} \cite{jax}; \texttt{jaxoplanet} \citep{jaxoplanet}; \texttt{numpyro} \citep{phan2019}; and ChainConsumer \citep{Hinton2016}.
}

\bibliography{bib}

\end{document}